\documentclass[%
reprint,
superscriptaddress,
altaffilletter,
showpacs,
showkeys,
amsmath, amssymb, aps, prl, floatfix,
]{revtex4-1}

\usepackage{graphicx}
\usepackage{dcolumn}
\usepackage{bm}
\usepackage{comment}
\usepackage[usenames,dvipsnames]{color}
\usepackage{hyperref}
\hypersetup{colorlinks = true, allcolors = blue}
\usepackage[mathlines]{lineno}
\usepackage{multirow}
\DeclareGraphicsExtensions{.pdf,.png,.jpg}

\newcommand{\Se}{$^{82}$Se}

\newcommand{\ZnSe}{Zn$^{82}$Se} 
\newcommand{\DBD}{0$\nu$DBD}

\newcommand{\THO}{$^{232}\mathrm{Th}$}

\newcommand{\cupidz}{CUPID-0}

\newcommand{\exposureZnSe}{16.59\,kg$\times$yr} %enriched only
\newcommand{\exposure}{8.82\,kg$\times$yr} %enriched only
\newcommand{\LimitFundamental}  {T$_{1/2}^{0\nu}$($^{82}$Se) $>$4.6$\times$10$^{24}$\,yr}
\newcommand{\Limitmbb} {m$_{\beta\beta} <$ (263 -- 545)\,meV}
\begin{document}

\title{Final Result on the Neutrinoless Double Beta Decay of $^{82}$Se with CUPID-0}

\newcommand{\sapienza}{\affiliation{Dipartimento di Fisica, Sapienza Universit\`a di Roma, P.le Aldo Moro 2, 00185, Roma, Italy}}
\newcommand{\infnroma}{\affiliation{INFN, Sezione di Roma, P.le Aldo Moro 2, 00185, Roma, Italy}}
\newcommand{\cnr}{\affiliation{Consiglio Nazionale delle Ricerche, Istituto di Nanotecnologia, c/o Dip. Fisica, Sapienza Università di Roma, 00185, Rome, Italy}}
\newcommand{\lnl}{\affiliation{INFN  Laboratori Nazionali di Legnaro, I-35020 Legnaro (Pd) - Italy}}
\newcommand{\lngs}{\affiliation{INFN  Laboratori Nazionali del Gran Sasso, I-67100 Assergi (AQ) - Italy}}
\newcommand{\lbl}{\affiliation{Lawrence Berkeley National Laboratory , Berkeley, California 94720, USA}}
\newcommand{\infnge}{\affiliation{INFN  Sezione di Genova, I-16146 Genova - Italy}}
\newcommand{\unige}{\affiliation{Dipartimento di Fisica, Universit\`{a} di Genova, I-16146 Genova - Italy}}
\newcommand{\infnmib}{\affiliation{INFN  Sezione di Milano - Bicocca, I-20126 Milano - Italy}}
\newcommand{\unimib}{\affiliation{Dipartimento di Fisica, Universit\`{a} di Milano - Bicocca, I-20126 Milano - Italy}}
\newcommand{\csnsm}{\affiliation{CNRS/CSNSM, Centre de Sciences Nucl$\acute{e}$aires et de Sciences de la Mati$\grave{e}$re, 91405 Orsay, France}}
\newcommand{\cea}{\affiliation{IRFU, CEA, Universit$\acute{e}$ Paris-Saclay, F-91191 Gif-sur-Yvette, France}}
\newcommand{\gssi}{\affiliation{Gran Sasso Science Institute, 67100, L'Aquila - Italy}}
\newcommand{\usc}{\affiliation{Department of Physics  and Astronomy, University of South Carolina, Columbia, SC 29208 - USA}}
\newcommand{\mpi}{\affiliation{Max-Planck-Institut für Physik, D-80805 München, Germany}}
\newcommand{\dis}{\affiliation{DISAT, Universit\`a dell'Insubria, 22100 Como, Italy}}

\author{O.~Azzolini}\lnl
\author{J.W.~Beeman}\lbl
\author{F.~Bellini}\sapienza\infnroma
\author{M.~Beretta}\altaffiliation{Present address: Department of Physics, University of California, Berkeley, CA 94720, USA}\unimib\infnmib
\author{M.~Biassoni}\infnmib
\author{C.~Brofferio}\unimib\infnmib
\author{C.~Bucci} \lngs
\author{S.~Capelli}\unimib\infnmib
\author{V.~Caracciolo}\altaffiliation{Present address: Dipartimento di Fisica, Universit\`{a} di Roma Tor Vergata, I-00133, Rome, Italy } \lngs
\author{L.~Cardani}\email[Corresponding author: ]{laura.cardani@roma1.infn.it}\infnroma
\author{P.~Carniti}\unimib\infnmib
\author{N.~Casali}\infnroma
\author{D.~Chiesa}\unimib\infnmib
\author{M.~Clemenza}\unimib\infnmib
\author{I.~Colantoni}\infnroma\cnr
\author{O.~Cremonesi}\infnmib
\author{A.~Cruciani}\infnroma
\author{A.~D'Addabbo} \lngs
\author{I.~Dafinei}\infnroma
\author{F.~De~Dominicis}\gssi\infnroma
\author{S.~Di~Domizio}\unige\infnge
\author{F.~Ferroni}\infnroma\gssi
\author{L.~Gironi}\unimib\infnmib
\author{A.~Giuliani}\csnsm
\author{P.~Gorla} \lngs
\author{C.~Gotti}\infnmib
\author{G.~Keppel}\lnl
\author{M.~Martinez}\altaffiliation{Present address: Centro de Astropart\'iculas y Física de Altas Energ\'ias, Universidad de
Zaragoza, and ARAID, Fundaci\'on Agencia Aragonesa para la Investigaci\'on y el Desarrollo, Gobierno de Arag\'on, Zaragoza 50018, Spain}\sapienza\infnroma
\author{S.~Nagorny}\altaffiliation{Present address: Department of Physics $\&$ Engineering Physics Astronomy, Queen's University
Kingston, Ontario, K7L 3N6 Kingston, Canada}\lngs
\author{M.~Nastasi}\unimib\infnmib
\author{S.~Nisi}\lngs
\author{C.~Nones}\cea
\author{D.~Orlandi}\lngs
\author{L.~Pagnanini}\lngs\gssi
\author{M.~Pallavicini}\unige\infnge
\author{L.~Pattavina}\altaffiliation{Present address: Physik-Department and Excellence Cluster Origins, Technische Universit{\"a}t M{\"u}nchen, 85747
Garching, Germany}\lngs
\author{M.~Pavan}\unimib\infnmib
\author{G.~Pessina}\infnmib
\author{V.~Pettinacci}\infnroma
\author{S.~Pirro}\lngs
\author{S.~Pozzi}\unimib\infnmib
\author{E.~Previtali}\unimib\lngs
\author{A.~Puiu}\lngs\gssi
\author{C.~Rusconi}\lngs\usc
\author{K.~Sch\"affner}\altaffiliation{Present address: Max-Planck-Institut f{\"u}r Physik, 80805 M{\"u}nchen - Germany}\lngs
\author{C.~Tomei}\infnroma
\author{M.~Vignati}\sapienza\infnroma
\author{A.~S.~Zolotarova}\cea

\date{\today}

\begin{abstract}
\cupidz, an array of \ZnSe\ cryogenic calorimeters, was the first medium-scale demonstrator of the scintillating bolometers technology. The first project phase (March 2017 -- December 2018) allowed the most stringent limit on the neutrinoless double beta decay half-life of the isotope of interest, \Se, to be set. After a six months long detector upgrade, \cupidz\ began its second and last phase (June 2019 -- February 2020). 
In this letter, we describe the search for neutrinoless double beta decay of \Se\ with a total exposure (phase I + II) of \exposure\ of isotope. We set a limit on the half-life of \Se\ to the ground state of  $^{82}$Kr of T$^{0\nu}_{1/2}$($^{82}$Se)$>$ 4.6$\times \mathrm{10}^{24}$ yr (90\% credible interval), corresponding to an effective Majorana neutrino mass \Limitmbb. 
We also set the most stringent lower limits on the neutrinoless decays of \Se\ to the  0$_1^+$, 2$_1^+$ and 2$_2^+$ excited states of $^{82}$Kr, finding  
1.8$\times$10$^{23}$\,yr, 
3.0$\times$10$^{23}$\,yr, 
3.2$\times$10$^{23}$\,yr  (90$\%$ credible interval) respectively.

\end{abstract}

\pacs{}
\keywords{neutrinoless double beta decay, Zn$^{82}$Se scintillating cryogenic calorimeters}
\maketitle
%\linenumbers
The possibility of observing neutrinoless double beta decay~\cite{GoeppertMayer,Furry:1939qr} (\DBD) has been intriguing an increasing number of scientists, in part because its detection would be a unique probe of the nature of neutrinos~\cite{Doi:1985dx}, in part because it would demonstrate the existence of a process that violates a fundamental symmetry of the Standard Model of Particle Physics: ($B-L$), $B$ and $L$ being the baryon and lepton number, respectively. Such violation would have exciting consequences for theories trying to explain the excess of matter over anti-matter in the Universe~\cite{KUZMIN198536,FUKUGITA198645,PhysRevD.42.3344}.

The signature of \DBD\ is very clean: the two electrons emitted in the decay share the whole Q-value of the transition, i.e. the difference between the mass of the parent and daughter nuclei ($\sim$MeV for the majority of the isotopes for which the decay is possible). As a consequence, \DBD\ would appear as a monochromatic peak at the Q-value in the sum energy spectrum of the two electrons. 
The central challenge in building an experiment to detect \DBD\ is the extreme rarity of this process.
Current limits on its half-life exceed 10$^{25}$ or even 10$^{26}$ years, depending on the isotope of interest~\cite{Adams_2020,Agostini_2020,Gando_2016,Anton_2019}.

Future detectors must deploy at least 10$^{26}$-10$^{27}$ DBD emitters to be able to detect few signal events over years of data-taking. A convincing claim for the discovery of such a feeble signal relies on the possibility of suppressing the background level in the region of interest to zero. A superb energy resolution would also be a key asset to disentangle a potential \DBD\ signal from the background, including the tail of the naturally occurring two-neutrino double beta decay. With typical half-lives of 10$^{18}$ -- 10$^{21}$ yr, this process produces a continuous energy spectrum extending up to the Q-value, resulting in a potential background in the signal region~\cite{Barabash:2020nck}.

The increasing interest in \DBD\ led to the implementation of many detection technologies, spanning from gas~\cite{nextcollaboration2021sensitivity} or liquid~\cite{nexocollaboration2021nexo} time projection chambers to germanium diodes~\cite{legendcollaboration2021legend1000}, and  scintillators~\cite{gando2019neutrinoless,Albanese_2021,PhysRevD.98.092007}. 
Calorimeters operated at cryogenic temperatures around 10\,mK (historically also called bolometers~\cite{Fiorini:1984}) are among the leading technologies in the field.
In bolometers, the thermal signal produced by an energy deposit is converted into an electrical signal using sensors with strong dependency of the resistance on the temperature~\cite{Pirro:2017}.
The crystal used as the energy absorber can be grown out of compounds containing the emitter of interest for \DBD, allowing a very high efficiency on the containment of the two electrons emitted in the process to be reached. Furthermore, the readout of the calorimetric signal offers an exquisite energy resolution (better than 1$\%$ FWHM, see e.g. Refs~\cite{CUORE:2021mvw,Armengaud_2021,Azzolini:2018tum}). Finally, the possibility of deploying a large detector mass was proved by the CUORE experiment, that recently surpassed the tonne$\cdot$yr exposure, setting the most stringent limit on the half-life of the $^{130}$Te \DBD~\cite{Artusa:2014lgv,CUORE:2021mvw}.

The CUPID collaboration (CUORE Upgrade  with Particle IDentification~\cite{CUPIDInterestGroup:2019inu}) is designing  a bolometric experiment to reach a \DBD\ half-life sensitivity of 1.5$\times$10$^{27}$\,yr.
This ambitious goal requires novel technological approaches allowing the suppression of the CUORE background index by two orders of magnitude. 
Starting from the CUORE background model, we inferred that the dominant background component are $\alpha$ particles produced by contamination in the material constituting the detector itself~\cite{Alduino_2017}. The main advance of CUPID compared to CUORE is using scintillating bolometers~\cite{Pirro:2005ar}, i.e. crystals emitting light at cryogenic temperatures. The simultaneous readout of the calorimetric signal and of the scintillation light enables particle identification and, thus, the rejection of the $\alpha$ background. 
Moreover, by choosing an isotope with a Q-value well above the 2.6\,MeV line of $^{208}$Tl (considered as the end-point of the natural $\gamma$ radioactivity), the CUPID signal lies in a region of the energy spectrum that is significantly less affected by background induced by $\beta$ interactions.

\cupidz\ is the first medium-scale detector based on this technology. Building on the experience of the LUCIFER project~\cite{Beeman:lucifer:2013,Beeman:2013vda,Beeman:2015xjv,Artusa:2016maw},
the \cupidz\ collaboration chose \Se\ as the emitter of interest (Q-value: 2997.9$\pm$0.3 keV~\cite{Lincoln_2013}), to be embedded in enriched ZnSe bolometers. This isotope was also investigated by the NEMO-3 collaboration, which set a 90$\%$ C.L. lower limit on its \DBD\ half-life of 2.5$\times$10$^{23}$\,yr (for the decay to the ground state~\cite{Arnold:2018tmo}) and 2.3$\times$10$^{22}$\,yr (for the 0$^+_1$ case~\cite{Arnold_2020}).
Following the novel procedure described in Ref.~\cite{Dafinei:2017xpc}, the \cupidz\ collaboration grew 24 Zn$^{82}$Se crystals 95$\%$ enriched in $^{82}$Se (total mass of 9.65 kg, corresponding to 5.13 kg of $^{82}$Se) and two natural ZnSe crystals (total mass of 0.85 kg, corresponding to 40\,g of $^{82}$Se). 
The crystals were arranged in five closely-packed towers using a copper mechanical structure and polytetrafluoroethylene clamps. 
The ZnSe crystals were interleaved by light detectors (LDs), consisting of a 170-$\mu$m-thick Ge disk~\cite{Beeman:2013zva} coated with a 60-nm-thick SiO layer to increase the light collection efficiency~\cite{Mancuso:2014paa}.
Both the ZnSe crystals and  light detectors were equipped with a thermal sensor, a neutron transmutation doped Ge thermistor (NTD~\cite{Haller}) and with a Si Joule heater~\cite{ANDREOTTI2012161} to enable the offline correction of pulse amplitude variations due to small thermal drifts~\cite{Alessandrello:1998,Alfonso:2017wzp}. The voltage variations across the NTDs were continuously saved on disk using an 18 bit analog-to-digital converter operating at 1\,kHz for the Zn$^{82}$Se and 2\,kHz for the (faster) LD~\cite{DiDomizio:2018ldc}, after being amplified and filtered with a six-pole anti-aliasing active Bessel filter (120 dB/decade)~\cite{Arnaboldi:2017aek,Arnaboldi:2010zz}.
The detector was operated at 10\,mK using an Oxford 1000 $^3$He/$^4$He dilution refrigerator located in Hall A of the Laboratori Nazionali del Gran Sasso. The major upgrade of the refrigerator, that was previosly used by the Cuoricino and CUORE-0 collaborations, consisted in the installation of a two-stage vibration damping system similar to the one described in Ref.~\cite{Pirro:2006mu}.
More details on the detector design, construction and commissioning can be found in Ref.~\cite{Azzolini:2018tum}.

In the first project phase, each crystal was surrounded by a 3M Vikuiti reflective foil. The second project phase was carried out without reflector, as the large light output of ZnSe at cryogenic temperatures already ensured an excellent particle identification capability.
On the other hand, the reflecting foil absorbed $\alpha$ particles produced by decays on the crystal surface, thus limiting our capability of reconstructing the correct topology of surface events~\cite{Azzolini:2019nmi,chiesa:2021}.

The study presented in this letter exploits the statistics collected in both phase-I and phase-II to search for the $^{82}$Se \DBD\ to the ground and excited states of its daughter, $^{82}$Kr. 
Today, the highest sensitivity on these processes was obtained by the phase-I of the \cupidz\ experiment, that reached 90$\%$ credible interval (C.I.) lower limit on the \DBD\ half-life of 3.5$\times$10$^{24}$\,yr~\cite{Azzolini:2018dyb,Azzolini:2019tta}). The half-lives of the decays to the first excited states were bound to T$_{1/2}^{0\nu}$($^{82}$Se $\rightarrow ^{82}$Kr$_{0_1^+}$) $>$8.1$\times$10$^{22}$\,yr,
T$_{1/2}^{0\nu}$($^{82}$Se $\rightarrow ^{82}$Kr$_{2_1^+}$)$>$1.1$\times$10$^{23}$\,yr,
and 
T$_{1/2}^{0\nu}$($^{82}$Se $\rightarrow ^{82}$Kr$_{2_2^+}$)$>$8.4$\times$10$^{22}$\,yr~\cite{Azzolini:2018oph}.
In this work, we improved these results by using the full statistics of \cupidz.

We define the experimental signatures associated to the \DBD\ to the ground state and to the excited states of $^{82}$Kr as follows. In the decay to the ground state, we expect the two emitted electrons to be fully contained in a single ZnSe crystal (single-crystal event). In the decay to the excited states, additional de-excitation $\gamma$ rays are produced (see Supplemental Material for a pictorial view of the decay scheme) and, while electrons are generally fully contained in the crystal where the decay took place, $\gamma$'s have a high escape probability and can be totally or partly absorbed by other ZnSe crystals (multiple-crystal event). 

Table~\ref{Table:signatures} summarises the signatures of interest for the ground and excited states decays and their containment efficiency, evaluated by a Monte Carlo simulation accounting also for the angular correlation of the emitted $\gamma$'s. More details about the \cupidz\ implementation of the (GEANT-4 based~\cite{AGOSTINELLI2003250,1610988,ALLISON2016186}) Monte Carlo simulation can be found in Ref. \cite{Azzolini:2019nmi}.
To improve the signal-to-background ratio, we focused only on signatures in which (i) the electrons were fully contained in the crystal, (ii) at least one of the emitted photons deposited its full energy inside a single crystal. We also discarded signatures in which the containment efficiency was smaller than 0.01$\%$, as they do not improve the sensitivity of the analysis. \\

\begin{table}[thb]
\centering
\caption{Decay configurations and corresponding event topology of $^{82}$Se to the ground and excited states of $^{82}$Kr, involving one, two or three ZnSe crystals; $\beta\beta$ represents the electrons emitted in the decay to the ground state ($\beta\beta_0$), to the 2$_1^+$ state ($\beta\beta_1$), 2$_2^+$ state  ($\beta\beta_2$) and 0$_1^+$  state ($\beta\beta_3$);  $\gamma_i$ are the emitted photons. In the \emph{Topology} column, the vertical bars separate energy deposits in a first (E$_{main}$), second (E$_{coinc}^{I}$) and third (E$_{coinc}^{II}$) crystal. The containment efficiency of each decay scheme is reported in column $\epsilon_{i}^{cont}$ (the subscript $i$ refers to the signature number, reported in the first column). In the last column we labelled the signatures used in this analysis; processes resulting in the same signature are labelled with the same letter (note that two signatures cannot be disentangled due to the finite energy resolution of the detector and are thus labelled B$_1$ and B$_2$). Decay schemes with a containment efficiency lower than 0.1$\%$ were not taken into account.}

\begin{tabular}{llccccc}
&Topology                                                                      &E$_{main}$         &E$_{coinc}^{I}$          &E$_{coinc}^{II}$      &$\epsilon_{i}^{cont}$		     &\\
&                    								&[keV]			&[keV]			&[keV]                       &[$\%$] 			     	  &\\

\hline
\hline
\vspace{0.03 cm}

1 &$\beta\beta_1$ $\mid$ $\gamma_1$                           &2221.4 		&776.5				&-				&1.817$\pm$0.009           &A\\     
\hline
\hline
2 &$\beta\beta_2$ $\mid$ $\gamma_1$                           &1523.0 		&776.5				&-				&0.604$\pm$0.004  		&B$_2$\\     
3 &$\beta\beta_2$ $\mid$ $\gamma_2$                           &1523.0 		&698.4				&-				&0.664$\pm$0.004 	 	&F\\     
4 &$\beta\beta_2$ $\mid$ $\gamma_3$                           &1523.0 		&1474.9				&-				&0.919$\pm$0.007 		 &G\\     
5 &$\beta\beta_2$ $\mid$ $\gamma_1$ +  $\gamma_2$ &1523.0 		&1474.9				&-				&0.0141$\pm$0.0004 	 &G\\     
6 &$\beta\beta_2$ + $\gamma_1$ $\mid$ $\gamma_2$  &2299.5 		&698.4				&-				&0.201$\pm$0.002  		&E\\     
7 &$\beta\beta_2$ + $\gamma_2$ $\mid$ $\gamma_1$  &2221.4  		&776.5				&-				&0.211$\pm$0.002 		 &A\\     
8 &$\beta\beta_2$ $\mid$ $\gamma_1$ $\mid$ $\gamma_2$  &1523.0  		&776.5				&698.4		&$<$0.01 		                 &-\\     
\hline
\hline
\vspace{0.03 cm}

9 &$\beta\beta_3$ $\mid$ $\gamma_1$                           &1510.3 		&776.5				&-				&0.606$\pm$0.006 		&B$_1$\\     
10 &$\beta\beta_3$ $\mid$ $\gamma_4$                           &1510.3 		&711.1			    	&-				&0.660$\pm$0.006  		&D\\     
11 &$\beta\beta_3$ + $\gamma_1$ $\mid$ $\gamma_4$  &2286.8 		&711.1 				&-				&0.196$\pm$0.003 		&C\\     
12 &$\beta\beta_3$ + $\gamma_4$ $\mid$ $\gamma_1$  &2221.4 		&776.5				&-				&0.200$\pm$0.003  		&A\\     
13 &$\beta\beta_3$ $\mid$ $\gamma_1$ + $\gamma_4$  &1510.3 		&1487.6				&-				&0.0146$\pm$0.0009  	&-\\     
14 &$\beta\beta_3$ $\mid$ $\gamma_1$ $\mid$ $\gamma_4$  &1510.3 		&776.5			&711.1			&$<$0.01  		        &-\\     
\hline
\hline
\vspace{0.03 cm}
15 &$\beta\beta_0$ $\mid$ ---						&2997.9 		&-					&-				&81.0$\pm$0.2                &ground\\

\end{tabular}
\label{Table:signatures}
\end{table}

The data processing followed closely the strategy of \cupidz\ phase-I, whose comprehensive description can be found in Ref.~\cite{Azzolini:2018yye}. 
Physics data of phase-I and phase-II were divided in 13 datasets (each lasting between 1 and 2 months). At the beginning and at the end of each dataset, we performed a $\sim$4 days long calibration with $^{232}$Th sources placed out of the cryostat.

Pulses recorded by the ZnSe crystals and their LDs were processed using the Optimum Filter technique~\cite{Gatti:1986cw}. 
The filtered amplitudes of heat pulses were first corrected by small temperature drifts of the cryostat by using the Si Joule heater~\cite{Alfonso:2017wzp}, and then converted into energy by using the periodic $\gamma$ calibrations with \THO\ sources. To convert amplitudes into energy values, we used second-degree polynomial functions with zero intercept.
Finally, the energy resolution of the ZnSe crystals was improved by removing the correlation with the corresponding light signals~\cite{Beretta_2019}.
In contrast to other bolometric detectors~\cite{CUORE:2016acf,Armatol_2021,Schmidt_2020,Armengaud_2017}, in \cupidz\ the FWHM energy resolution shows a linear dependency on the energy (see Supplemental Material for more details and data). 
At the Q-value of the $^{82}$Se \DBD\ we obtain 21.8$\pm$0.3 keV FWHM. In the energy region for the search of the \DBD\ to the excited states (energies of interest in Table~\ref{Table:signatures}) it ranges from 8.9$\pm$0.1 to 17.9$\pm$0.2 keV FWHM.

The data selection comprised two steps. The first one applies basic selection cuts on the quality of bolometric pulses. The second step is especially designed for background rejection and thus is optimised separately for the study of the decay to the ground and excited states.

In the first selection procedure, we run a software derivative trigger on the continuously acquired data-stream~\cite{DiDomizio:2018ldc}.
We used the (flagged) pulses injected by the heaters in each ZnSe crystal to quantify the trigger efficiency at different energies. Each trigger efficiency was multiplied by the ``energy reconstruction" efficiency. To compute this value, we injected heater pulses with the same energy all over the dataset. We then processed the heater pulses through the same analysis chain as particle pulses and fit the heater peak with a Gaussian function. The energy reconstruction efficiency was defined as the ratio of the events reconstructed within 3 sigma from the mean and the total number of injected heater pulses. The obtained value was stable during the entire life of the experiment and equal to (99.2$\pm$0.5)$\%$.
We then rejected time periods in which the cryostat was unstable (for example, after liquid helium refills, or during earthquakes) reducing the live-time by about 1$\%$.

The triggered events were selected using a series of parameters: the value and slope of the baseline before the trigger (a proxy for the temperature value and temperature stability immediately before the pulse occurs), the pulse rise and decay time, the number of triggers in the same window, and some shape-dependent parameters derived by the Optimum Filter~\cite{Azzolini:2018yye}.
Such parameters allowed us to disentangle signal candidates from electronics spikes, crystal thermal contractions, pulses affected by pile-up and very noisy events.
We characterised and corrected for the energy dependency of each parameter in order to guarantee energy-independent cuts (and thus energy-independent efficiency). The values of the cut were optimised for each parameter using the odd events belonging to the physics peak of $^{65}$Zn: we defined the ratio $r = \epsilon_S/\sqrt{\epsilon_B}$ ($\epsilon_S$ and $\epsilon_B$ being the signal and background efficiency respectively), and increased the cut value until $r$ reached a plateau.
We then used the even events belonging to the same peak to evaluate the selection efficiency and obtained $\epsilon_{selection} = (92.3 \pm 0.7)\%$. This value was cross-checked at other energies using the $^{40}$K and $^{208}$Tl peaks and obtained consistent results.

\begin{figure}[thb]
\centerline{\includegraphics[width=8.5cm]{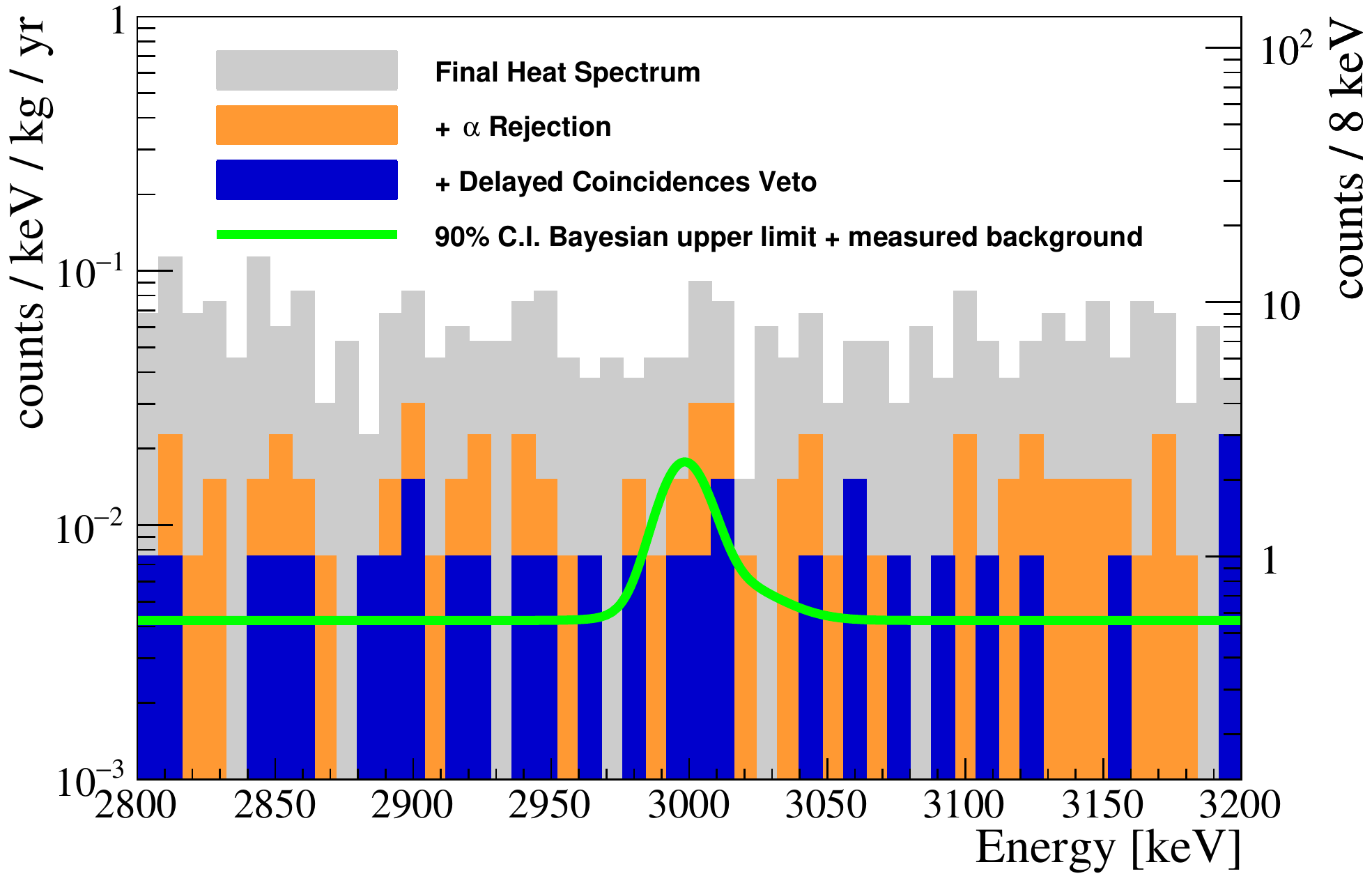}}
\caption{Physics spectrum with a $^{82}$Se exposure of \exposure\ (\exposureZnSe\ in ZnSe). Grey: spectrum of pulses selected according to their bolometric quality, with the additional condition that a single ZnSe detector triggered the event. Orange: same spectrum after rejecting $\alpha$ events. Blue: events surviving the delayed coincidence veto with potential $^{212}$Bi parents. Green line: 90$\%$ C.I. Bayesian upper limit (see text) superimposed to the measured background.}
\label{fig:background}
\end{figure}

After performing a selection on the quality of bolometric pulses, the analysis followed two different paths for the decay to the ground and excited states.
Concerning the decay to the ground state, we adopted the strategy outlined in Refs~\cite{Azzolini:2018dyb,Azzolini:2019tta,Azzolini:2018yye}. We selected only events in which a single ZnSe crystal triggered, as Table~\ref{Table:signatures} shows that in the vast majority of cases the two electrons are contained within the crystal where the decay occurred. 
Figure~\ref{fig:background} shows the spectrum obtained using the 22 enriched crystals with better performance, i.e. discarding two enriched crystals that have shown a poor bolometric performance due to a different growth procedure. 
The total $^{82}$Se exposure is  \exposure, corresponding to \exposureZnSe\ in ZnSe (see Supplemental Material for the spectrum in a wider energy range).

Since in the region of interest we expected a dominant background contribution stemming from $\alpha$ particles, we exploited the particle identification capability offered by scintillating bolometers to further select the events. In \cupidz, the particle identification is done exploiting the shape of the light pulses. The scintillation light emitted by $\beta/\gamma$ events, indeed, has a significantly slower time-development compared to the scintillation light emitted by $\alpha$ particles~\cite{Artusa:2016maw}. Furthermore, the time-development of scintillation pulses does not depend on the detector. We defined a parameter very sensitive to the shape of light pulses~\cite{Azzolini:2018yye} and we applied a cut procedure using such parameter (common to all detectors) in order to preserve the 98$\%$ signal efficiency while suppressing the $\alpha$ background to a negligible level (the probability for an $\alpha$ particle to survive this selection criterion is smaller than 10$^{-7}$). The resulting spectrum is reported in Figure~\ref{fig:background} - orange.

Finally, we applied a time veto to reject high-energy $\beta/\gamma$ events emitted by $^{208}$Tl decays. This isotope can be tagged by searching for the signature of its parent,  $^{212}$Bi. The $\alpha$ particle emitted by a $^{212}$Bi decay can be identified by \cupidz\ even at low energy (in case it loses a fraction of energy before interacting in the crystal). 
Exploiting the short half-life of $^{208}$Tl (3.05\,min) we rejected potential interactions due to this isotope by vetoing all events occurring within 7 half-lives after the detection of an $\alpha$ particle. This further data selection, allowed us to obtain the energy spectrum shown in Figure~\ref{fig:background} - blue.

To estimate the number of \DBD\ signals and background events in the region of interest, we performed a simultaneous un-binned extended maximum likelihood (UEML) fit. The signal was modelled with the detector response function (a bi-Gaussian line shape, as explained in the Supplemental Material), with position fixed at the Q-value and dataset-dependent resolution.
In addition, the signal efficiency was evaluated on a dataset basis and treated as a dataset-specific parameter in the fit. Averaging over the exposure of the 13 datasets, it resulted in (69.2$\pm$1.2)$\%$. This value comprises: 
the trigger and energy reconstruction efficiency (99.2$\%$), the selection efficiency of $\beta/\gamma$ events against $\alpha$ events (98$\%$), the selection of high-quality bolometric pulses (92.3$\%$ that,  combined with the delayed coincidence veto, diminished  to 87.9$\%$), and the containment efficiency (81.0$\%$).

The background was added as a flat component with different values for phase-I and phase-II. To include systematic uncertainties on energy scale, detector response function, efficiency and exposure, we weighted the likelihood with a Gaussian probability density function (fixing the mean and RMS to the mean value and uncertainty of each nuisance parameter). After a numerical integration of the likelihood, we obtained a 90$\%$ C.I. Bayesian upper limit on the decay width $\Gamma^{0\nu}$: 1.5$\times$10$^{-25}$\,yr$^{-1}$ corresponding to a lower limit on the half-life \LimitFundamental.
This result is slightly worse than the experimental median sensitivity of 7.0$\times$10$^{24}$\,yr 90$\%$ C.I.. However, the discrepancy can be explained by the statistical fluctuations of the background level that, unfortunately, presented an over fluctuation in the region of interest (Fig.~\ref{fig:background}). Such statistical fluctuations dominate the global uncertainty on the limit, while other systematic uncertainties can be considered negligible.

The resulting background index is (3.5$\pm$1.0)$\times$10$^{-3}$\,counts/keV/kg/yr in phase-I and (5.5$\pm$1.5)$\times$10$^{-3}$\,counts/keV/kg/yr in phase-II. 
Using the most updated values of the phase space factor~\cite{Kotila:2012zza,Stoica:2013lka} and nuclear matrix elements~\cite{Engel:2016xgb,Barea:2015kwa,Yao:2014uta,Rodriguez:2010mn,Menendez:2008jp,Simkovic:2018hiq}
we converted the limit on T$_{1/2}$ into a lower limit on the neutrino Majorana mass of \Limitmbb, today the most competitive result on $^{82}$Se.\\

The analysis of the $^{82}$Se decay to the excited states proceeded via a slightly simplified path. 
We first made a basic selection on the quality of bolometric pulses, as done in the previous analysis. Then, we required two crystals in the array to trigger the event in time-coincidence, as all the searched signatures are expected to produce simultaneous energy deposits in two detectors (Table~\ref{Table:signatures}). The energy of the primary channel was fixed to be within $\pm$3$\sigma$ of the nominal value E$_{coinc}^I$, resulting in an almost background free region without the need for more aggressive data selection techniques (see Supplemental Material for the figures of the seven spectra).

We made a simultaneous UEML fit to the seven spectra using the three values of $\Gamma^{0\nu}_{0_1^+}$, $\Gamma^{0\nu}_{2_2^+}$ and $\Gamma^{0\nu}_{2_1^+}$ as free parameters.
Following this approach, the number of signal events was defined for each signature as:
\begin{subequations}
\begin{align}
N_A^{sig}&=\xi\cdot\Big(\varepsilon_{1}\Gamma^{0\nu}_{2_1^+}+\varepsilon_7\Gamma^{0\nu}_{2_2^+}+\varepsilon_{12}\Gamma^{0\nu}_{0_1^+}\Big)\\
N_{B1}^{sig}&=\xi\cdot\varepsilon_{9}\Gamma^{0\nu}_{0_1^+} \\
N_{B2}^{sig}&=\xi\cdot\varepsilon_{2}\Gamma^{0\nu}_{2_2^+}\\
N_C^{sig}&=\xi\cdot\varepsilon_{11}\Gamma^{0\nu}_{0_1^+} \\
N_D^{sig}&=\xi\cdot\varepsilon_{10}\Gamma^{0\nu}_{0_1^+} \\
N_E^{sig}&=\xi\cdot\varepsilon_{6}\Gamma^{0\nu}_{2_2^+} \\
N_F^{sig}&=\xi\cdot\varepsilon_{3}\Gamma^{0\nu}_{2_2^+} \\
N_G^{sig}&=\xi\cdot\big(\varepsilon_{4}+\varepsilon_{5}\big)\Gamma^{0\nu}_{2_2^+}
\end{align}
\end{subequations}
where $\xi$ is the \cupidz\ total exposure and $\varepsilon_i$ is the product of the data selection efficiency by the containment efficiencies $\epsilon_i^{cont}$ reported in Table~\ref{Table:signatures}. 
The signal was modelled using the bi-Gaussian line shape, with position fixed at the nominal values of E$_{main}$ and FWHM fixed at the values obtained for each signature (see Supplemental Material for data).
The background was described using a flat component and, for signatures B, D, F and G, a peaking component due to the presence of the $^{40}$K peak in the proximity of the expected signal.
We included the systematic uncertainties already outlined in the analysis of the $^{82}$Se decay to the ground state of $^{82}$Kr.

With 90$\%$ C.I. median sensitivities of T$_{1/2}$($^{82}$Se $\rightarrow$ $^{82}$Kr$_{0_1^+}$) = 1.6$\times$10$^{23}$, 
T$_{1/2}$($^{82}$Se $\rightarrow$ $^{82}$Kr$_{2_1^+}$) = 2.9$\times$10$^{23}$ and
T$_{1/2}$($^{82}$Se $\rightarrow$ $^{82}$Kr$_{2_2^+}$) = 3.1$\times$10$^{23}$, 
we obtained the following limits on the partial half-lives:
\begin{subequations}
\begin{equation*}
T_{1/2}(^{82}\mathrm{Se} \rightarrow ^{82}\mathrm{Kr}_{0_1^+}) >1.8\times10^{23}\ \text{yr} 
\end{equation*}
\begin{equation*}
T_{1/2}(^{82}\mathrm{Se} \rightarrow ^{82}\mathrm{Kr}_{2_1^+}) >3.0\times10^{23}\ \text{yr} 
\end{equation*}
\begin{equation*}
T_{1/2}(^{82}\mathrm{Se} \rightarrow ^{82}\mathrm{Kr}_{2_2^+}) >3.2\times10^{23}\ \text{yr}
\end{equation*}
\end{subequations}
improving the existing limits by factor 2.2 to 3.8 depending on the signature (compared to the previous \cupidz\ limits).\\

This work was partially supported by the European Research Council (FP7/2007-2013) under Low-background Underground Cryogenic Installation For Elusive Rates Contract No. 247115. We are particularly grateful to M. Iannone for the help in all the stages of the detector construction, A. Pelosi for the construction of the assembly line, M. Guetti for the assistance in the cryogenic operations, R. Gaigher for the calibration system mechanics, M. Lindozzi for the development of cryostat monitoring system, M. Perego for his invaluable help,  the mechanical workshop of LNGS (E. Tatananni, A. Rotilio, A. Corsi, and B. Romualdi) for the continuous help in the overall setup design. We acknowledge the Dark Side Collaboration for the use of the low-radon clean room.
This work makes use of the DIANA data analysis and APOLLO data acquisition software which has been developed by the CUORICINO, CUORE, LUCIFER, and CUPID-0 Collaborations.

\end{document}